\documentclass[twocolumn, superscriptaddress]{revtex4}
\usepackage{txfonts}
\usepackage[dvipdfmx]{graphicx}
\usepackage[dvipdfmx]{color}
\usepackage{color}

\begin{document}

\title{Emergence of topological Hall effect in half-metallic manganite thin films by tuning perpendicular magnetic anisotropy}
\author{M. Nakamura}
\email{masao.nakamura@riken.jp}
\affiliation{RIKEN Center for Emergent Matter Science (CEMS), Wako, 351-0198, Japan}
\affiliation{PRESTO, Japan Science and Technology Agency (JST), Kawaguchi, 332-0012, Japan}
\author{D. Morikawa}
\affiliation{RIKEN Center for Emergent Matter Science (CEMS), Wako, 351-0198, Japan}
\author{X. Z. Yu}
\affiliation{RIKEN Center for Emergent Matter Science (CEMS), Wako, 351-0198, Japan}
\author{F. Kagawa}
\affiliation{RIKEN Center for Emergent Matter Science (CEMS), Wako, 351-0198, Japan}
\author{T. Arima}
\affiliation{RIKEN Center for Emergent Matter Science (CEMS), Wako, 351-0198, Japan}
\affiliation{Department of Advanced Materials Science, University of Tokyo, Kashiwa 277-8561, Japan}
\author{Y. Tokura}
\affiliation{RIKEN Center for Emergent Matter Science (CEMS), Wako, 351-0198, Japan}
\affiliation{Department of Applied Physics and Quantum Phase Electronics
Center (QPEC), University of Tokyo, Tokyo 113-8656, Japan} 
\author{M. Kawasaki}
\affiliation{RIKEN Center for Emergent Matter Science (CEMS), Wako, 351-0198, Japan}
\affiliation{Department of Applied Physics and Quantum Phase Electronics
Center (QPEC), University of Tokyo, Tokyo 113-8656, Japan} 
\date{\today}

\begin{abstract}
Magnetic materials hosting topological spin textures like magnetic skyrmion exhibit nontrivial Hall effect, namely, topological Hall effect (THE).  In this study, we demonstrate the emergence of THE in thin films of half-metallic perovskite manganites. To stabilize magnetic skyrmions, we control the perpendicular magnetic anisotropy by imposing a compressive epitaxial strain as well as by introducing a small Ru doping.  When the perpendicular magnetic anisotropy is tuned so that it is balanced with the magnetic dipolar interaction, the film exhibits a sizable THE in a magnetization reversal process. Real-space observations indicate the formation of skyrmions and some of them have high topological charge number. The present result opens up the possibility for novel functionalities that emerge under keen competition between the skyrmion phase and other rich phases of perovskite manganites with various orders in spin, charge, and orbital degrees of freedom.
\end{abstract}

\maketitle
\section{Introduction}
Perovskite manganites with a composition of $R_{1-x}$$A_{x}$MnO$_{3}$ ($R$ stands for a rare-earth ion, $A$ an alkaline earth ion, $x$ the band filling) as illustrated in Fig.~1(a) exhibit a wide variety of ordered structures in spin, charge, and orbital degrees of freedom depending on the band filling and the band width~\cite{Tokura2006}. Various unique phenomena observed in these systems  originate from the strong correlation between these electron degrees of freedom. A well-known example for one of such phenomena is the colossal magnetoresistance, which occurs due to the phase transition between a charge-orbital-ordered antiferromagnetic insulating state and a double-exchange ferromagnetic metallic state induced by a magnetic field. The conduction electron in the ferromagnetic-metallic state has a nearly 100~\% spin polarization, and hence it is recognized as a representative half-metal~\cite{Park1998}. Another example is the multiferroicity which appears in Mott insulator phase at $x = 0$. In this phase, the spontaneous electric polarization is induced by a non-collinear spiral order of the local spins, leading to the emergence of the non-trivial electromagnetic responses, such as the large modulation of the magnetization by an electric field~\cite{Cheong2007,Tokura2010}. 

\begin{figure}[b]
\begin{center}
\includegraphics[width=8cm]{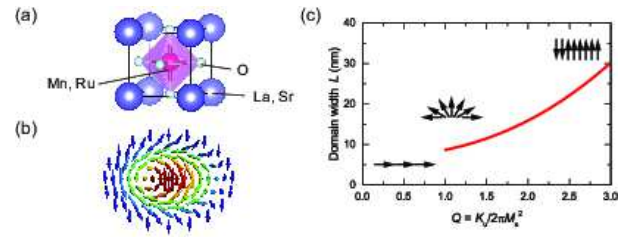}
\caption{(a) A schematic of the crystal structure of perovskite manganite. (b) A schematic spin texture of a single skyrmionic bubble. (c) Stripe domain width in a ferromagnetic thin film as a function of $Q$ factor calculated using Eq.~1. The parameters we used for the calculation are described in the main text. The inset shows the evolution of the characteristic spin configuration with $Q$ factor; spins pointing lateral direction with large domain ($Q < 1$), twisted configuration ($Q\sim1$), and spins pointing perpendicular direction with large domain ($Q > 1$).}
\end{center}
\end{figure}

When a spin-polarized conduction electron passes through a non-collinear local spin order, non-trivial electromagnetic coupling emerges, as typified by the topological Hall effect (THE) observed in the compounds hosting magnetic skyrmion~\cite{Bruno2004,Binz2008}. This Hall effect originates from the Berry phase acquired by the conduction electron whose spin is aligned to the local spin by the Hund's-rule coupling. The half-metallic manganite is an attractive system to examine the THE because its magnitude is proportional to the spin polarization. The magnetic skyrmion is a particle-like object with a whirling spin texture as illustrated in Fig.~1(b). The exact definition of the magnetic skyrmion is a solitonic state stabilized by a competition between the exchange interaction enforcing a parallel spin alignment and the Dzyaloshinskii-Moriya interaction (DMI) twisting the parallel spin alignment. Hence, it is observed in noncentrosymmetric magnets (chiral skyrmion)~\cite{Bogdanov1994,Resler2006,Muhlbauer2009,Yu2010}. Similar but slightly different particle-like magnetic texture is observed in centrosymmetric magnets known as the magnetic bubble which is stabilized by the magnetic dipolar interaction instead of DMI~\cite{Malozemoff,Hubert}. Although the chiral skyrmion and magnetic bubble are different in the magnetization profile in the core region~\cite{Kiselev2011}, their dynamical response and THE can be quantified by a common topological invariant called the skyrmion number ($N$), which is defined as a number of spheres wrapped by the constituent spins~\cite{Moutafis2009,Buttner2015,Nagaosa2013}. Hence, a magnetic bubble can be also regarded as a skyrmion in a broad sense, and is called `skyrmionic bubble'. Although the study on the skyrmionic bubble has a long history, it attracts recently a renewed interest , in particular, due to the richer skyrmion textures which are brought about by the underlying helicity degree of freedom~\cite{Nagaosa2013,Ezawa2010}. 
Furthermore, the skyrmionic bubbles can form only in single-phase bulk
crystals~\cite{Nagai2012,Nagao2013,Yu2013,Morikawa2015,Kotani2016,Yu2012,Wang2016}, but also in thin film multilayers, sometimes in conjunction with DMI~\cite{Finazzi2013,Li2014,Jiang2015,Gilbert2015,Woo2016,Lee2016,Soumyanarayanan2016}. The wide variety of materials choice is a major advantage of the skyrmionic bubble from the application point of view.

Although there are several reports on the observation of skyrmionic bubbles in bulk crystals of
manganites~\cite{Nagai2012,Nagao2013,Yu2013,Morikawa2015,Kotani2016}, neither the control of the skyrmion size nor the observation of THE has been achieved yet. Thin film structure provides an excellent platform for such a study. The size of the skyrmionic bubble in a thin film crucially depends on the magnitude of the uniaxial magnetic anisotropy. The quality factor $Q = K_{\mathrm{u}}/\Omega$, where $K_{\mathrm{u}}$ is the uniaxial magnetic anisotropy energy and $\Omega=2\pi M_{\mathrm{s}}^{2}$ ($M_{\mathrm{s}}$ is saturation magnetization) is the dipolar interaction energy, is known to be a good measure of the domain size~\cite{Yafet1988,Vedmedenko2000}. The domain size can be formed only when $Q\geq1$, and the domain width ($L$) becomes  larger with $Q$ as shown in Fig.~1(c). $L$ is calculated based on the following analytical solution,
\begin{equation}
L=\frac{5nJ\pi^{2}}{6\Omega_{\mathrm{L}}}\frac{\exp\left(\sqrt{nJ\pi^{4}(Q-1)\Omega_{\mathrm{S}}/\Omega_{\mathrm{L}}^{2}}+1\right)}{\sqrt{nJ\pi^{4}(Q-1)\Omega_{\mathrm{S}}/\Omega_{\mathrm{L}}^{2}}+1},
\end{equation}
where the film has $n$ layers ($n=t/a$, $t$ is the film thickness and $a$ the lattice constant), $J$ the exchange interaction energy,  $\Omega_{\mathrm{L}}=2\pi(nM_{\mathrm{s}})^2$, and $\Omega_{\mathrm{S}}=2\pi nM_{\mathrm{s}}^2$~\cite{Wu2004,Won2005}. The $Q$ dependence of the domain width shown in Fig.~1(c) is estimated using  $J=$2.5~meV~\cite{Moussa2007}, $n$~=~75 ($t$~=~30~nm) and $M_{\mathrm{s}}$~=~2.5~$\mu_{\mathrm{B}}$/f.u.. In this study, we tune the $Q$ value in thin films of manganite by controlling $K_{\mathrm{u}}$ using both the single-ion anisotropy induced by Ru doping and the epitaxial strain. We find that large THE appears when the perpendicular magnetic anisotropy and dipolar interaction are in keen competition.

\begin{figure}[htbp]
\begin{center}
\includegraphics[width=8cm]{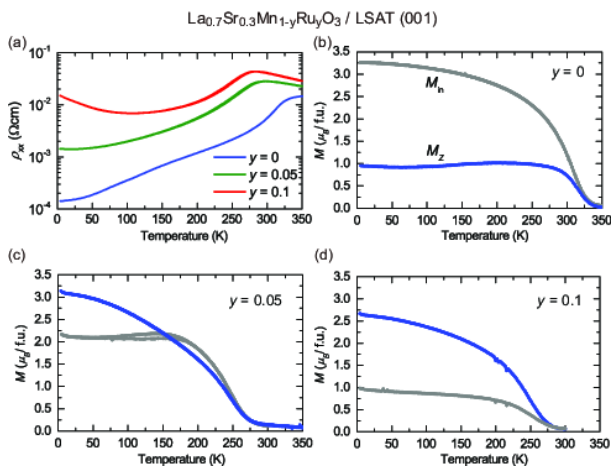}
\caption{(a) Temperature dependence of the resistivity for La$_{0.7}$Sr$_{0.3}$Mn$_{1-y}$Ru$_{y}$O$_{3}$ films on LSAT(001) substrate with $y = 0$, 0.05, and 0.1. (b)(c)(d) Temperature dependence of the magnetization for the same samples measured in a magnetic field of 0.1~T applied parallel ($M_{\mathrm{in}}$) and perpendicular ($M_{z}$) to the film surface.}
\end{center}
\end{figure}

\section{Sample Preparation}
Thin films of La$_{0.7}$Sr$_{0.3}$Mn$_{1-y}$Ru$_{y}$O$_{3}$ (LSMRO) with a thickness of 30~nm were grown on (001) surface of (LaAlO$_{3}$)$_{0.3}$(SrAl$_{0.5}$Ta$_{0.5}$O$_{3}$)$_{0.7}$ (LSAT) substrates by a pulsed laser deposition technique. The Ru concentration $y$ was varied from 0 to 0.1. As reported in Ref.~\cite{Yamada2005}, a too high growth temperature ($T_{\mathrm{growth}}$) or a low oxygen pressure ($P_{\mathrm{O}_{2}}$) causes the deficiency of Ru. We optimized the growth condition  at $T_{\mathrm{growth}}$ = 720~$^{\circ}$C and $P_{\mathrm{O}_{2}}$ = 40~mTorr to obtain
thin films with a stoichiometric composition and an atomically flat surface. We verified by x-ray diffraction measurements that the films grown under the optimum condition have pseudomorphic structures with their $c$-axis being elongated due to the compressive strain imposed by LSAT substrate whose lattice constant is smaller than that of LSMRO~\cite{Sahu2000}.

\section{Magnetic and Transport Properties}
Figure~2 displays temperature dependence of the resistivity and magnetization for the films with $y$ = 0, 0.05, and 0.1. All films show a metallic behavior concomitant with the appearance of the ferromagnetic state, whereas the magnetic easy axis changes its direction. At $y$ = 0, the magnetic easy axis lies along the in-plane direction ($M_{\mathrm{in}}$). As increasing the Ru concentration, $M_{\mathrm{in}}$ decreases and alternatively out-of-plane component ($M_{z}$) increases. $M_{\mathrm{in}}$ and $M_{\mathrm{z}}$ are comparable at $y=0.05$, and  $M_{z}$ dominates $y=0.1$. The switching of the magnetic easy axis with Ru concentration is also apparent in the magnetic-field dependence of the magnetization shown in Figs.~3(a), (b), and (c).
We thus  that the film with $y=0.05$ is in a state of $Q\sim1$, the ideal condition for the generation of small skyrmions.
To confirm the effect of the epitaxial strain on the magnetic anisotropy, we fabricated a thin film with $y$ = 0.05 on a SrTiO$_{3}$ (STO) substrate to apply a tensile strain. 
As shown in Fig.~4, the $M$-$H$ curve of the film grown on STO substrate indicates a robust in-plane magnetic anisotropy with a large coercive magnetic field, being consistent with the result in Ref.~\cite{Yamada2005}.

\begin{figure}[htbp]
\begin{center}
\includegraphics[width=8cm]{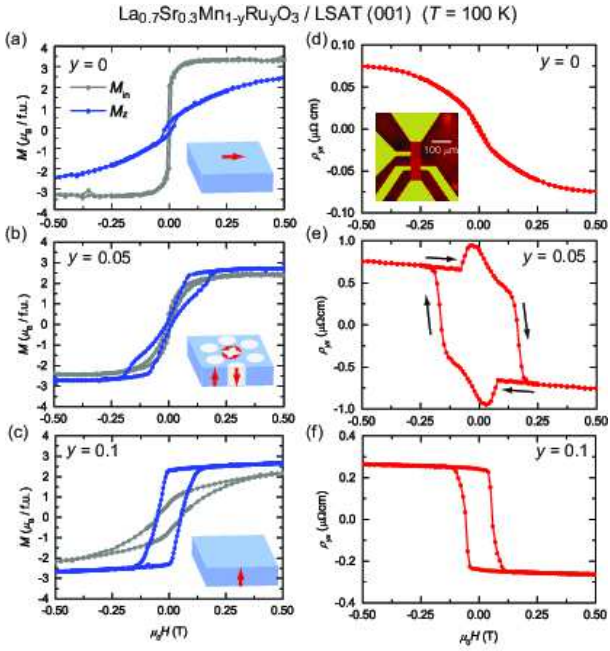}
\caption{(a)(b)(c) Magnetic-field ($H$) dependence of magnetization measured at a temperature of 100~K for La$_{0.7}$Sr$_{0.3}$Mn$_{1-y}$Ru$_{y}$O$_{3}$ (LSMRO) films grown on LSAT(001) substrates. $M_{\mathrm{in}}$ ($M_{z}$) denotes the magnetization measured by applying $H$ parallel (perpendicular) to the film surface. The spin texture expected from the magnetization curves at each Ru concentration is shown in the inset. (d)(e)(f) $H$ dependence of the Hall resistivity ($\rho_{yx}$) measured at 100~K. The inset of Fig.~3(g) is a picture of the Hall-bar device.}
\end{center}
\end{figure}

It has been known that the magnetic anisotropy in manganite films can be controlled by the epitaxial strain~\cite{Kwon1997,Wu1999,Dho2003}.  However, the realization of a perpendicularly magnetized state only by the epitaxial strain requires a sizable compressive strain imposed by a largely lattice-mismatched substrate. This, however, causes a partial strain relaxation and homogeneous magnetic state. Furthermore, it is difficult to realize continuous variation of the magnetic anisotropy due to the limited choice of the substrates. Therefore, there has been no report on the dense and small skyrmion formation in manganite films. Our results indicate that the combination of the compressive strain and Ru doping enables the perpendicularly magnetized state by a modest strain and a continuous control of the magnetic anisotropy by changing Ru concentration. The compressive biaxial epitaxial strain imposed by LSAT substrate lifts the $t_{\mathrm{2g}}$ orbital degeneracy of the doped Ru$^{4+}$ ion with $xy$ orbital being at the lowest energy. The restored orbital angular momentum in a heavy element Ru induces the large single-ion anisotropy necessary for the perpendicular magnetic anisotropy. 

Figures~3(d)-(f) show the Hall resistivity ($\rho_{yx}$) of these films in Figs.~3(d)-(f). We used photolithography and ion-milling to pattern the films in Hall-bar shapes (inset of Fig.~3(d)). The $\rho_{yx}$ of the films with $y = 0$ and 0.1 are almost proportional to $M_{z}$. On the contrary, the film with $y$ = 0.05 shows a large contribution of an additional Hall signal in the small magnetic field region.  In conventional magnets, the ordinary Hall resistivity ($\rho_{yx}^{\mathrm{O}}$) proportional to the magnetic field and the anomalous Hall resistivity ($\rho_{yx}^{\mathrm{A}}$) proportional to $M_{z}$ contribute to $\rho_{yx}$. In compounds hosting non-zero spin chirality such as skyrmion, another contribution to $\rho_{yx}$ emerges, that is THE. Since the film with $y = 0.05$ is located near the critical point of $Q = 1$, the skyrmion density is expected to be enhanced compared to the other compositions. We thus consider that the additional Hall signal in this compound stems from THE.

\begin{figure}[htbp]
\begin{center}
\includegraphics[width=8cm]{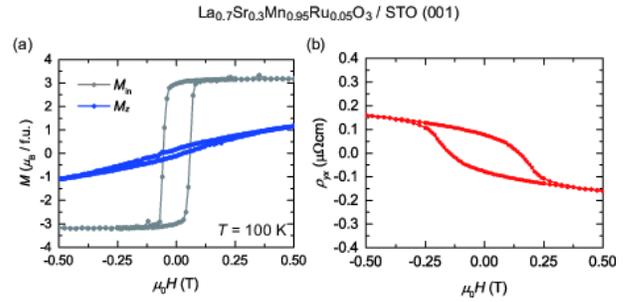}
\caption{(a) Magnetic-field dependence of the magnetization of LSMRO($y = 0.05$)/STO(001) film measured at 100~K. (b) Magnetic-field dependence of the Hall resistivity for the same film measured at 100~K.}
\end{center}
\end{figure}

\begin{figure}[bp]
\begin{center}
\includegraphics[width=8cm]{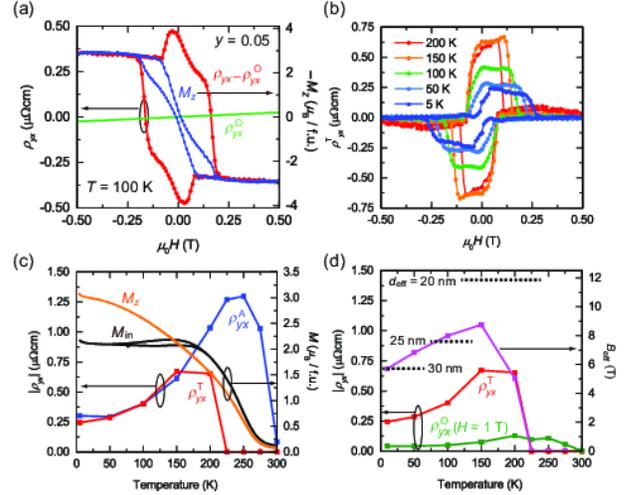}
\caption{(a) $H$ dependence of $\rho_{yx}$, ordinary Hall resistivity ($\rho_{yx}^{\mathrm{O}}$), and $M_{z}$ for LSMRO film with $y$ = 0.05 at 100~K. $\rho_{yx}^{\mathrm{O}}$ was derived by linear fitting of $\rho_{yx}$-$H$ curve above 1~T of magnetic field. The scale of the vertical axis for $\rho_{yx}$ (left axis) and that for $M_{z}$ is adjusted so that they overlap at 0.5~T. (b) Temperature evolution of topological Hall resistivity ($\rho_{yx}^{\mathrm{T}}$) versus $H$. (c) Temperature dependence of $\rho_{yx}^{\mathrm{T}}$ and $\rho_{yx}^{\mathrm{A}}$ (left axis) and $M_{\mathrm{in}}$ and $M_{z}$. (right axis). $M_{\mathrm{in}}$ and $M_{z}$ are measured at a magnetic field of 0.1~T. (d) Temperature dependence of $\rho_{yx}^{\mathrm{T}}$ and $\rho_{yx}^{\mathrm{O}}$ at $H$ = 1~T as well as the effective field ($B_{\mathrm{eff}}$) which is derived by $\rho_{yx}^{\mathrm{T}}/\rho_{yx}^{\mathrm{O}}$($H$ = 1~T). The relation between $B_{\mathrm{eff}}$ and effective skyrmion diameter ($d_{\mathrm{eff}}$) is shown for several representative values of $d_{\mathrm{eff}}$. }
\end{center}
\end{figure}

Figure~5(a) shows $H$-dependence of $\rho_{yx}$, $\rho_{yx}^{\mathrm{O}}$, and $M_{z}$ for the film with $y=0.05$ at 100~K. $\rho_{yx}^{\mathrm{O}}$ was derived from the linear fitting of $\rho_{yx}$ measured at magnetic fields above 1~T where $\rho_{yx}$ shows a linear dependence on $H$. The contribution of $\rho_{yx}^{\mathrm{A}}$ is derived by fitting $\rho_{yx}-\rho_{yx}^{\mathrm{O}}$ with $\alpha M_{z}$ ($\alpha$ is a constant) in $H$ outside of the hysteresis. The topological Hall resistivity ($\rho_{yx}^{\mathrm{T}}$) was derived by subtracting $\rho_{yx}^{\mathrm{O}}$ and $\rho_{yx}^{\mathrm{A}}$ from the total $\rho_{yx}$, \textit{i.e.}, $\rho_{yx}^{\mathrm{T}} = \rho_{yx}-\rho_{yx}^{\mathrm{O}}-\rho_{yx}^{\mathrm{A}}$~\cite{Lee2009,Neubauer2009,Kanazawa2011,Huang2012,Porter2014,Yokouchi2014}. The derivation of $\rho_{yx}^{\mathrm{T}}$ was carried out by the same procedure at other temperatures, and $\rho_{yx}^{\mathrm{T}}$ traces are shown in Fig.~5(b). Figure~5(c) plots the peak value of $\rho_{yx}^{\mathrm{T}}$ as well as $M_{\mathrm{in}}$, $M_{z}$, and $\rho_{yx}^{\mathrm{A}}$ as a function of temperature. $\rho_{yx}^{\mathrm{A}}$ appears at the ferromagnetic transition temperature ($T_{\mathrm{C}}=280$~K) and monotonously decreases with lowering temperature. By contrast, $\rho_{yx}^{\mathrm{T}}$ appears at 200~K which is apparently lower than $T_{\mathrm{C}}$, and has a peak at 150~K. The temperature dependence of $\rho_{yx}^{\mathrm{T}}$ is related to the variation of the magnetic anisotropy. The temperature dependence of the magnetization indicates that $M_{\mathrm{in}}$ and $M_{z}$ becomes comparable at around 150~K, and $M_{\mathrm{in}}$ ($M_{z}$) dominates above (below) this temperature. Above 150~K, $\rho_{yx}^{\mathrm{T}}$ is absent because the magnetic easy axis lies in-plane ($Q < 1$) and the skyrmion is not formed. As the system approaches the spin reorientation temperature, $\rho_{yx}^{\mathrm{T}}$ appears and reaches the maximum when $Q\sim1$ state is realized. At this point, the skyrmion density is expected to be largest. At lower temperatures, the perpendicular magnetic anisotropy is further enhanced, which causes the increase of $Q$, leading to the reduction of the skyrmion density and accordingly the magnitude of $\rho_{yx}^{\mathrm{T}}$.

Additional evidence for the skyrmion formation in the film with $y=0.05$ is found in the dependence of the THE on the magnetic field direction~\cite{Yokouchi2014}. Figure~6(a) shows that THE vanishes with increasing the tilt angle of the magnetic field ($\theta$), whereas the anomalous Hall effect remains almost constant. The reduction of $\rho_{yx}^{\mathrm{T}}$ under inclined field is defined by $\Delta\rho_{yx}^{\mathrm{T}}(\theta)=\rho_{yx}(\theta)-\rho_{yx}(\theta=30^{\circ})$, and $\Delta\rho_{yx}^{\mathrm{T}}(\theta)/\Delta\rho_{yx}^{\mathrm{T}}(\theta=0^{\circ})$ is plotted as a function of $\theta$ in Fig.~6(b). $\rho_{yx}^{\mathrm{T}}$ reduces gradually, but it rapidly drops at around $\theta = 5^{\circ}\sim10^{\circ}$.
The skyrmion diameter ($d$) is estimated as $d=t/\sin\theta_{\mathrm{s}}$ ($\theta_{\mathrm{s}}$ is the angle where THE disappears and $t$ is the film thickness)~\cite{Ohuchi2015}. The top axis of Fig.~6(b) shows the scale of $t/\sin\theta$. It indicates that the typical value of $d$ is about 200-300~nm.

\begin{figure}[tbp]
\begin{center}
\includegraphics[width=8cm]{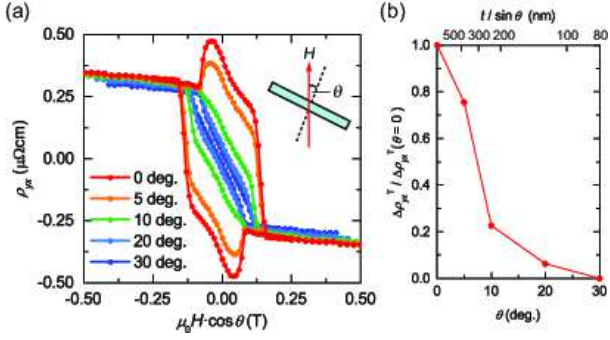}
\caption{(a) $\rho_{yx}$-$H$ curves for LSMRO ($y = 0.05$) film at 100~K measured in magnetic fields with various inclination angle ($\theta$). Data are plotted as a function of the perpendicular component of magnetic field ($\mu_{0}H\cos\theta$). The inset depicts the relation between $\theta$ and $H$. (b) The reduction of the $\rho_{yx}^{\mathrm{T}}$ under inclined field is defined by $\Delta\rho_{yx}^{\mathrm{T}}(\theta) = \rho_{yx}(\theta)-\rho_{yx}(\theta=30^{\circ})$, and $\Delta\rho_{yx}^{\mathrm{T}}(\theta)/\Delta\rho_{yx}^{\mathrm{T}}(\theta=0^{\circ})$ is plotted as a function of $\theta$. The top axis shows a scale of $t/\sin\theta$ ($t$ = 40~nm).}
\end{center}
\end{figure}

The size of skyrmion can be also estimated from the magnitude of THE. A single skyrmion gives an effective field of the magnetic flux quantum $\Phi_{0}=h/e$ to a conduction electron, where $h$ is the Planck's constant and $e$ is the elementary charge. Therefore, $\rho_{yx}^{\mathrm{T}}$ and the skyrmion density $\Phi$ is related as $\rho_{yx}^{\mathrm{T}}=PR_{0}\Phi_{0}\Phi$~\cite{Neubauer2009}, where $R_{0}$ is the ordinary Hall coefficient and $P$ is the spin polarization. $P\sim 1$ in the metallic state of the perovskite manganites~\cite{Park1998}. We derive the effective field ($B_{\mathrm{eff}}\Phi_{0}\Phi$) by comparing the temperature dependence of $\rho_{yx}^{\mathrm{T}}$ and $\rho_{yx}^{\mathrm{O}}$ shown in Fig.~5(d). 
The density of skyrmion estimated from $B_{\mathrm{eff}}$ is about 2000~$\mu\mathrm{m}^{-2}$ at 100~K (1300~$\mu\mathrm{m}^{-2}$ at 10~K), which means that the effective diameter of skyrmion ($d_{\mathrm{eff}}$) is 24~nm (30~nm) assuming a close-packed hexagonal lattice. This value is several times smaller than that estimated from the angle dependence of THE. We shall discuss the possible origins of this discrepancy later.

\begin{figure*}[htbp]
\begin{center}
\includegraphics[width=13cm]{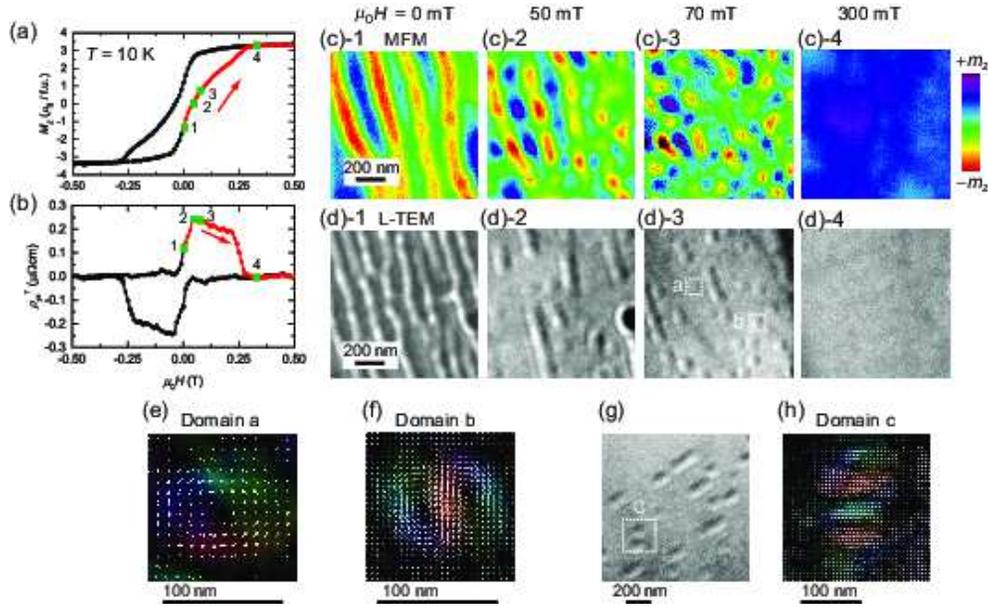}
\caption{(a) $M_{z}$-$H$  and (b)  $\rho_{yx}^{\mathrm{T}}$-$H$ curves for LSMRO($y = 0.05$) film measured at 10~K. Data points shown in red indicate the scan direction traced during the real-space observations. The green dots are data points at which real-space magnetic domain images shown in Figs. 7(c) are recorded. (c) Magnetic-force microscope (MFM) images taken at 10~K and (d) Lorentz transmission microscope (L-TEM) images taken at 100~K at various magnetic fields; $\mu_{0}H$ = 0~T (7(c)-1 and 7(d)-1), 50~mT (7(c)-2 and 7(d)-2), 70~mT (7(c)-3 and 7(d)-3), and 300~mT (7(c)-4 and 7(d)-4). (e)-(g) Result of the transport-of-intensity equation (TIE) analyses for L-TEM images. Figures~7(e), 7(f) are TIE analyses for domain-a, domain-b denoted in Fig.~7(d)-3, respectively. The domain-a is a single skyrmion with skyrmion number $N = 1$ and the domain-b is a biskyrmion with $N = 2$. Figure~7(g) is a L-TEM image taken at different area of the sample. Figure~7(h) is TIE analysis for domain-c denoted in Fig.~7(g), which indicates a chain of 4 skyrmions.}
\end{center}
\end{figure*}

\section{Real-Space Observations}
We now describe the real-space observation of spin textures using magnetic-force microscopy (MFM) and Lorentz transmission electron microscopy (L-TEM). These two techniques offer complementary information on the spin texture; MFM is sensitive to the out-of-plane magnetization component, whereas L-TEM can detect the in-plane component~\cite{Yu2010,Milde2013,Soumyanarayanan2016}. 
MFM observation was performed using an Attocube low-temperature atomic-force platform (AttoAFMI). During the scan, we detected the resonant frequency shift of the cantilever, which originates from the interaction between the magnetization of the tip and stray magnetic field from the film. The frequency shift is proportional to the second derivative of the local magnetic field with respect to $z$ direction. We employed a Nanosensors PPP-MFMR cantilever. The magnetization direction of the tip was defined by applying $-$5~T in $z$ direction at 10~K. Then, magnetic field was turned back to zero and MFM images were recorded at positive magnetic fields. We did not observe the magnetization reversal of the tip up to +300~mT. The excitation amplitude of the cantilever was 5~ nm. Before the MFM observation, we recorded the surface topography and stored the tilt of the sample. Then, the MFM images were taken under constant height (100~nm) mode. It was not necessary to correct the MFM data by the topography signal because the sample is atomically flat.

Figures~7(c) show the MFM images taken at four representative magnetic fields. The corresponding $H$ dependence of $M_{z}$ and $\rho_{yx}^{\mathrm{T}}$ are shown in Figs.~7(a) and 7(b). More detailed $H$ dependence of MFM image is displayed in Fig.~8. The MFM image at zero field shows a stripe domain (Fig.~7(c)-1). By applying a w, modulation patterns appear inside of the stripe domain as seen in the image at 50~mT (Fig.~7(c)-2). They are pinched off and become discrete domains as seen in the image at 70~mT (Fig.~7(c)-3). Near this magnetic field, $\rho_{yx}^{\mathrm{T}}$  reaches its maximum value. The domain structure finally disappears when $H$ exceeds the field value necessary to saturate $M_{z}$ as seen in the image at 300~mT (Fig.~7(c)-4), and accordingly, $\rho_{yx}^{\mathrm{T}}$ vanishes.

\begin{figure}[tbp]
\begin{center}
\includegraphics[width=6.5cm]{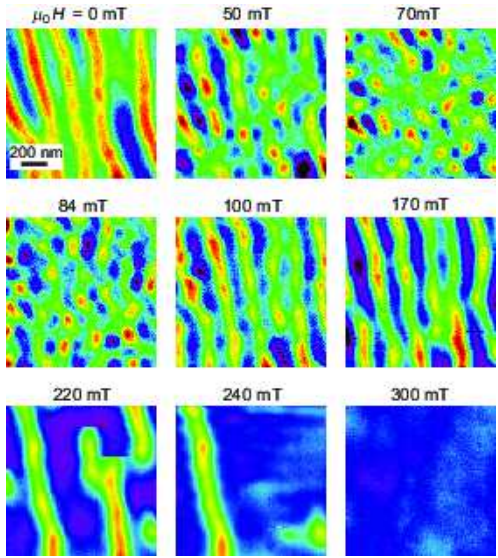}
\caption{Magnetic-field dependence of MFM images for the LSMRO($y = 0.05$)/LSAT(001) film taken at 10~K. }
\end{center}
\end{figure}

To clarify a possible existence of the periodic structure in the skyrmion phase as well as the stripe phase, the autocorrelation function analysis was performed for the MFM image at 70~mT as displayed in Fig.~9. Average distance between the adjacent skyrmions derived from the autocorrelation function is 300~nm (390~nm) along the red (blue) line (Fig.~9(c)). The skyrmion diameter estimated from the MFM image spread between 90 and 200~nm, and the average size is about 150~nm.

L-TEM observation was performed at the Lorentz TEM mode of a conventional transmission electron microscope (JEM-2100F (JEOL)).
The sample for L-TEM observation was prepared by thinning the substrate of the film. 
The substrate was firstly thinned by mechanical polishing, and then, a part of the substrate was further thinned by Ar ion milling at a low temperature. 
Since the compressive epitaxial strain from LSAT substrate is important to induce the perpendicular magnetic anisotropy, the ion milling was stopped leaving the substrate thickness of about 200~nm. The magnetic field was controlled by changing the objective lens current. The magnetization textures were obtained by analyzing defocused L-TEM images with a software Qpt based on transport-of-intensity-equation (TIE)~\cite{Ishizuka2005}.

The L-TEM images shown in Figs.~7(d) exhibit a similar evolution of the domain structure in the magnetic field as observed with MFM; stripe domain at zero field, isolated circular domain at 70~mT, and single domain state at 300~mT. The typical diameter of the isolated domain is about 100~nm. The size and density of magnetic textures observed by L-TEM are slightly different from those observed by MFM, probably caused by the partial strain relaxation in L-TEM sample which arose during the thinning process of the substrate. Another reason can be a different measurement temperature. The TIE analysis for a discrete domain denoted by domain-a in Fig.~7(d)-3 indicates that at the center the magnetic moments are pointing normal to the film surface and the outside moments have a swirling structure (Fig.~7(e)). This is a characteristic spin texture of the skyrmion with a skyrmion number $N = 1$. Furthermore, the TIE analysis for another discrete domain denoted by domain-b reveals a bound state of two skyrmions with opposite helicities, namely the biskyrmion state (Fig.~7(f))~\cite{Yu2013,Wang2016}, which has $N = 2$ and therefore contributes doubly to the THE. We also found a domain in which four skyrmions are bound and possibly characterized by $N = 4$ as shown in Figs.~7(g) and 7(h). 

\begin{figure}[tbp]
\begin{center}
\includegraphics[width=7cm]{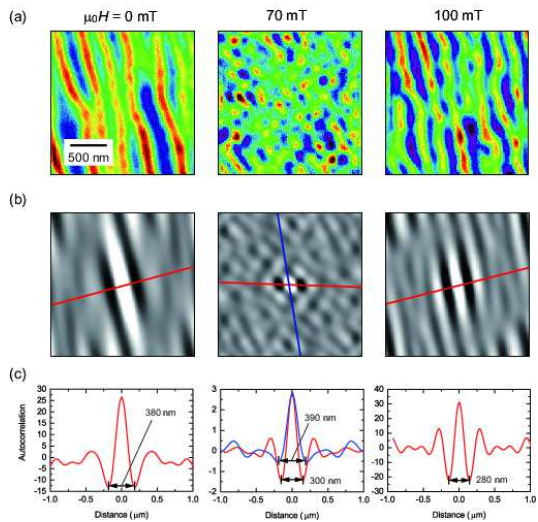}
\caption{(a) MFM images for the  LSMRO($y = 0.05$)/LSAT(001) film taken at three different magnetic fields. (b) Two-dimensional autocorrelation function for the MFM images shown in Fig.~9(a). (c) Line profiles of the autocorrelation function along the lines in Fig.~9(b).}
\end{center}
\end{figure}

The diameter of skyrmion ($d$) observed by MFM and L-TEM is in the range of 100-200~nm, which is consistent with that estimated from the magnetic-field angle dependence of the THE, but $d_{\mathrm{eff}}$ estimated from the magnitude of THE is much smaller and is in the range of 20-30~nm. The fact of such discrepancy in skyrmion size indicates that THE is enhanced by several times by some reason. One apparent reason is the existence of skyrmions having multiple topological charges as revealed by the L-TEM images. Another possible reason is the THE in the momentum space as observed in frustrated and disordered ferromagnets~\cite{Taguchi2001,Lyanda-Geller2001}. In the latter case, Hall effect is usually observed only near $T_{\mathrm{C}}$ in perovskite manganites associated with thermally-driven hedgehog spin configuration~\cite{Lyanda-Geller2001}. In our film with $Q\sim1$, the solid angle of the nearest-neighbor local moments can remain finite even at low temperatures due to a keen competition between the exchange and the dipolar interactions, which may induce the non-trivial Hall effect. 

\section{Conclusions}
In conclusion, we controlled the magnetic anisotropy in thin films of a half-metallic perovskite manganite. 
We found the emergence of large THE in the film with the perpendicular magnetic anisotropy being balanced with the magnetic dipolar interaction. MFM and L-TEM observations revealed the existence of a few hundreds nanometer sized skyrmion bubbles in the film.
We find several times enhancement in the magnitude of THE compared with that expected from real-space observations, 
indicating a possibility of other mechanisms to enhance THE in perovskite manganites.

\begin{acknowledgments}
We appreciate D. Shindo (Tohoku Univ. and RIKEN), T. Akashi (Hitachi Ltd.), and T. Tanigaki (Hitachi Ltd.) for their supporting in L-TEM observation. We also appreciate W. Koshibae (RIKEN) for fruitful discussions and D. Maryenko (RIKEN) for critical reading of our manuscript.  This work was partially supported by PRESTO JST (JPMJPR16R5).
\end{acknowledgments}

\end{document}